\documentclass[a4paper,twocolumn]{article}
\usepackage{epsf}
\normalsize

\def\bp{{\bf p}}
\def\bq{{\bf q}}
\def\bk{{\bf k}}

\def\Im{{\rm Im}}

\def\up{\uparrow}
\def\down{\downarrow}
\def\eps{\epsilon}
\def\gam{\gamma}
\def\om{\omega}
\def\sg{\sigma}
\def\Sg{\Sigma}
\def\sgn{{\rm sgn\:}}
\makeatletter
\def\@abstract{} \def\abstract#1{\gdef\@abstract{#1}}
\def\@address{} \def\address#1{\gdef\@address{#1}}
\def\@pacs{} \def\pacs#1{\gdef\@pacs{#1}}
\renewcommand{\@maketitle}{%
  \newpage
  \null
  \vskip 2em%
  \begin{center}%
  \let \footnote \thanks
    {\Large\bfseries \@title \par}%
    \vskip 1.5em%
    {\large
      \lineskip .5em%
      \begin{tabular}[t]{c}%
        \@author\\
	{\small\em \@address}
      \end{tabular}\par}%
    \vskip 1em%
    {\small \@date}%
    \ifx\@abstract\@empty\else%
	\vskip .5em \begin{minipage}{14cm}%
	\small\@abstract\\[1ex]%
	\ifx\@pacs\@empty\else\mbox{{\bfseries PACS:}~\@pacs\fi}%
	\end{minipage}\par%
    \fi
  \end{center}%
  \par
  \vskip 1.5em}
\makeatother
\begin{document}
\title{Spectral Function of 2D Fermi Liquids}
\author{Christoph J.\ Halboth and Walter Metzner}
\address{Sektion Physik, Universit\"at M\"unchen,
Theresienstra{\ss}e 37, D-80333 M\"unchen, Germany}
\date{\today}
\abstract{%
We show that the spectral function for single-particle excitations in a
two-dimensional Fermi liquid has Lorentzian shape in the low energy limit.
Landau quasi-particles have a uniquely defined spectral weight and a decay
rate which is much smaller than the quasi-particle energy. By contrast,
perturbation theory and the T-matrix approximation yield spurious
deviations from Fermi liquid behavior, which are particularly pronounced
for a linearized dispersion relation.
}
\pacs{05.30.Fk,~71.10-w,~71.10.Ay}
\maketitle
\section*{\normalsize{1.~Introduction}}
The low energy physics of interacting Fermi systems is usually governed
by Landau's Fermi liquid theory, as long as no binding or symmetry-breaking
occurs \cite{LAN}.
Fermi liquid theory is based on the existence of quasi-particles, i.e.\
fermionic single-particle excitations with an energy-momentum relation 
$\xi_{\bk}$ that vanishes linearly as $\bk$ approaches the Fermi surface
$\cal F$ of the system.
In three-dimensional Fermi liquids quasi-particles decay with a rate 
$\gam_{\bk}$ that vanishes quadratically near $\cal F$,
as a consequence of the restricted phase space for low-energy scattering
processes. The single-particle spectral function $\rho(\bk,\xi)$ exhibits
a Lorentzian-shaped peak of width $\gam_{\bk}$ as a function of the 
energy variable $\xi$ for $\bk$ close to $\cal F$. 

In one-dimensional Fermi systems Landau's quasi-particles are unstable:
the decay rate of single-particle excitations with sharp momenta is at 
least of the order of their energy and their spectral weight vanishes 
in the low-energy limit, giving rise to so-called Luttinger liquid
behavior \cite{VOI}. 

No consensus has so far been reached on the properties of single-particle
excitations in two-dimensional Fermi systems. 
The anomalous properties of electrons moving in the $\mathrm{CuO_2}$-planes of
high-$T_c$ superconductors have motivated numerous
speculations on a possible breakdown of Fermi liquid theory in 2D systems,
at least for sufficiently strong interaction strength \cite{MCD} and, 
according to Anderson \cite{AND}, even for weak coupling.

The development of new powerful renormalization group techniques 
\cite{FT,BG} has recently opened the way towards a rigorous
non-perturbative control of interacting Fermi systems for sufficiently
small yet finite coupling strength \cite{FMRT}. 
Significant rigorous results have already been derived for 2D systems.
For example, the existence of a finite jump in the momentum distribution
has been established for two-dimensional Fermi systems with a 
non-symmetry-broken ground state \cite{FKLT}. This excludes 
Luttinger liquid behavior in weakly interacting 2D systems.

So far, however, no rigorous results could be obtained for real-time
(or frequency) dynamic quantities such as spectral functions.
In particular, even at weak coupling the shape of the spectral function
for single-particle excitations and the existence of well-defined 
quasi-particles are still under debate.
Indeed, within second order perturbation theory the imaginary part of
the self-energy $\Sg(\bk,\xi)$ exhibits a sharp peak near 
$\xi = \xi_{\bk}$ in two dimensions \cite{CDM}, to be contrasted with
the simple quadratic and $\bk$-independent energy-dependence of 
$\Im\Sg(\bk,\xi)$ in 3D. 
Such a peak in the self-energy leads to a spectral function with two 
separate maxima instead of a single quasi-particle peak \cite{CDM,FO}.
However, analyzing an analytic continuation of one-dimensional Luttinger
liquids to higher dimensionality $d$, Castellani, Di Castro and one of
us \cite{CDM,MCD} have pointed out that such peaks in the perturbative 
self-energy are smeared out in a random phase approximation (RPA) and 
probably also in an exact solution.
Most recently a breakdown of Fermi liquid theory has been inferred 
from a divergence of the slope of the self-energy, computed within a
T-matrix approximation (TMA), at $\xi = \xi_{\bk}$ \cite{YF}.
\medskip

In this article we present explicit results for the self-energy
computed within perturbation theory, TMA and RPA for
a two-dimensional prototypical model with local interactions.
We will then provide several simple arguments showing that the RPA,
which yields a Fermi liquid type result for self-energy and spectral 
function, is qualitatively correct, while unrenormalized perturbation
theory and the T-matrix approximation produce artificial singularities
in 2D systems.
\medskip

\section*{\normalsize{2.~Perturbation theory and T-matrix}}

We consider an isotropic continuum model with a local coupling as a 
prototype for Fermi systems with short-range repulsive interactions,
\begin{equation}\label{eq1}
 H = \sum_{\bk,\sg} \eps_{\bk} \, a^{\dag}_{\bk\sg} a_{\bk\sg} 
 \, + \, {g \over V} \sum_{\bk,\bk',\bq}  
 a^{\dag}_{\bk+\bq\up} a_{\bk\up} \> 
 a^{\dag}_{\bk'-\bq\down} a_{\bk'\down} ,
\end{equation}
where $a^{\dag}_{\bk\sg}$ and $a_{\bk\sg}$ are the usual creation and
annihilation operators for spin-$1 \over 2$ fermions, $\eps_{\bk}$ is
an (isotropic) dispersion relation, $g\!>\!0$ a coupling constant, and $V$
the volume of the system. Local interactions between particles with
parallel spin do not contribute due to the Pauli exclusion principle.
A cutoff must be imposed for large momenta to make the model 
well-defined.

The spectral function for single-particle excitations $\rho(\bk,\xi)$
is directly related to the imaginary part of the one-particle Green 
function at real frequencies \cite{AGD}
\begin{equation}\label{eq2}
 G(\bk,\xi) = {1 \over \xi - (\eps_{\bk} - \mu) - \Sg(\bk,\xi)} .
\end{equation}
Here $\mu$ is the chemical potential (fixing the particle density).
Note that $\rho(\bk,\xi)$ can be measured by angular resolved
photoemission.

\medskip

\begin{figure}
\epsfbox{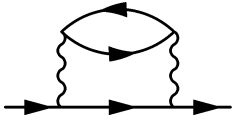}
\caption{Second order Feynman diagram contributing to $\Sg(\bk,\xi)$.}
\label{fig1}
\end{figure}
\begin{figure}
\epsfbox{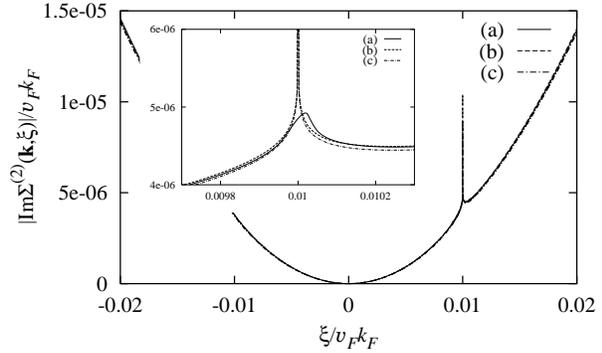}
\caption{Numerical results for $\Im\Sg^{(2)}(\bk,\xi)$ at fixed $|\bk|
= 1.01 k_F$ for (a) a quadratic and (b) a linear dispersion 
relation, compared to (c) the asymptotic analytic result (\ref{eq3}). 
The fine-structure of the peak near $\xi=\xi_{\bk}$ is shown in the
inset.}
\label{fig2}
\end{figure}

To first order in the coupling constant $g$, the self-energy is a 
constant which can be absorbed by shifting the chemical potential.
Non-trivial dynamics enters at second order, where for a local 
interaction only a single Feynman diagram (see Fig.\ \ref{fig1})
yields an 
energy-dependent contribution $\Sg^{(2)}(\bk,\xi)$. 
The imaginary part, $\Im\Sg^{(2)}(\bk,\xi)$, is cutoff independent and
can be expressed as a two-dimensional integral, which can be easily
performed numerically. 
Low-energy results for a quadratic (free-particle) dispersion 
$\eps_{\bk} = \bk^2/2m$
and a linear dispersion $\eps_{\bk} = v_F |\bk|$ are shown in
Fig.\ \ref{fig2}.
The Fermi momentum $k_F$ and the Fermi velocity $v_F$ has been set one 
in both cases ($v_F = k_F/m$ for quadratic dispersion),
and $\bk$ has been fixed slightly outside the Fermi surface, i.e.\ 
$|\bk| = 1.01 k_F$.
Note that the two curves almost coincide except very close to the point
$\xi = \xi_{\bk} = \eps_{\bk} \!-\! \mu$, while the structure of the peak
near $\xi_{\bk}$ depends sensitively on the curvature of $\eps_{\bk}$
near $k_F$.
Indeed, a strikingly simple analytic expression can be derived for the
asymptotic low-energy behavior of $\Im\Sg^{(2)}$ (see Appendix):
\begin{equation}\label{eq3}
\begin{array}{l}
\displaystyle
|\Im\Sg^{(2)}(\bk,\xi)| = g^2 \, {k_F \over (2\pi)^3 v_F^3} \,
\xi^2 \times \\
\displaystyle
\qquad\times \left[ 1 + \ln 4 \sqrt{3} - {3 \over 2} \ln|\xi/v_F k_F| - 
 {\kappa \over 2} - \ln|\kappa + 1| \right. \\
\displaystyle
\qquad\left. 
  - {1 \over 2} \ln|\kappa - 1|
  + {1 \over 4} (\kappa - 1)^2 
   \ln\left|{\kappa + 1 \over \kappa -1}\right| \right]
 \, + \, {\cal O}(|\xi|^{5/2}) \\
\end{array}
\end{equation}
for small $\xi$ at fixed $\kappa = \xi_{\bk}/\xi \neq 1$. 
For generic (non-linear) dispersion relations there are non-universal 
deviations from the leading asymptotic result in a tiny region of width
$\xi_{\bk}^2$ around $\xi = \xi_{\bk}$.
For $\bk = \bk_F$ (i.e.\ $\kappa = 0$) one recovers the known perturbative
result $\Im\Sg^{(2)}(\bk_F,\xi) \propto \xi^2 \log|\xi| + 
{\cal O}(|\xi|^2)$ \cite{FUJ}.
The logarithmic singularity for $\xi \sim \xi_{\bk}$ has already been
noted earlier in the literature \cite{CDM,FO}, but the complete 
asymptotic low-energy expression (\ref{eq3}) for the second order self-energy
has not been found so far.

As pointed out in another context by Castellani et al. \cite{CDM} and
also by Fukuyama and Ogata \cite{FO}, a peak in the self-energy as in
Fig.\ \ref{fig2} would lead to a spectral function with two maxima centered 
around the quasi-particle energy $\xi_{\bk}$, which is reminiscent of
the two-peak structure of spectral functions in one-dimensional 
Luttinger liquids \cite{MSV,VOI}.
\medskip

\begin{figure}
\epsfbox{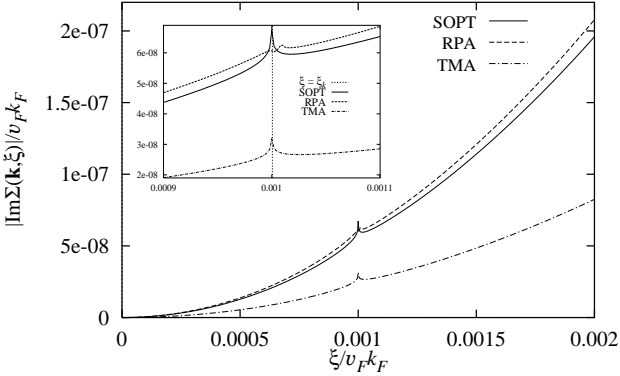}
\caption{Numerical results for $\Im\Sg(\bk,\xi)$ at fixed
 $|\bk| = 1.001k_F$ within second order perturbation theory, TMA and
 RPA, for a coupling $g = 1 {v_F \over k_F}$ and a quadratic
 $\epsilon_{\bk}$; the TMA requires a cutoff for large momenta which has
 been chosen as $\Lambda = 5k_F$.} 
\label{fig3}
\end{figure}

Several authors have computed the self-energy for two-dimensional
Fermi systems within the so-called T-matrix approximation (TMA), i.e.\ 
summing all particle-particle ladder diagrams \cite{BLO,FHN,ER}.
This approximation is expected to be asymptotically exact at low
density $n \to 0$ in dimensions $d \geq 2$, where corrections are smaller
by a factor $n$ in 3D but only by a factor $1/|\log(n)|$ in 2D \cite{ER}.
Numerical results for the imaginary part of the T-matrix self-energy
$\Sg^T(\bk,\xi)$, calculated for the model (\ref{eq1}), are shown in
Fig.\ \ref{fig3}.
The peak at $\xi\!=\!\xi_{\bk}$ is now bounded even for a linear $\eps_{\bk}$.
However, Yokoyama and Fukuyama \cite{YF} have recently found that the real
part of $\Sg^T(\bk,\xi)$ has infinite slope at $\xi = \xi_{\bk}$
for a linearized dispersion relation.
Evaluating the renormalization factor
\begin{equation}\label{eq4}
 Z(\bk,\xi) = \left[1-\partial\Sg(\bk,\xi)/\partial\xi \right]^{-1}
\end{equation}
in the special limit $\bk \to \bk_F$, $\xi \to 0$ with $\xi/\xi_{\bk}
\to 1$ they obtained the result "$Z \to 0$", which was interpreted as
a confirmation of Anderson's \cite{AND} ideas about a breakdown of Fermi
liquid theory in two dimensions.
This interpretation seems to be misleading.
Firstly, one cannot replace the function $Z(\bk,\xi)$ by a single number
"$Z$" if $Z(\bk,\xi)$ has no (unique) limit for $\bk \to \bk_F$,
$\xi \to 0$. Indeed, $Z(\bk,\xi)$ vanishes {\em only}\/ in the above 
special limit. 
In particular, the jump in the momentum distribution function across
the Fermi surface, which is also given by "$Z$" in a conventional Fermi
liquid, does {\em not}\/ vanish within the TMA in 2D systems!
Secondly, one cannot trust a singular result for a correlation function
obtained from a truncated asymptotic expansion with respect to a "small"
parameter, since higher order corrections are generally not 
{\em uniformly}\/ small for all momenta and energies. 
One should be even more worried in a case where corrections are 
suppressed at best by logarithmic factors $1/|\log(n)|$ \cite{EMH}.
Indeed, the artificial sensitivity of the peaks in $\Sg(\bk,\xi)$ on 
irrelevant non-linear terms in the dispersion relation found in perturbation
theory and TMA indicates that the exact self-energy will look differently.
\medskip

\section*{\normalsize{3.~Random phase approximation}}

A self-energy with a regular energy-dependence which does not depend 
sensitively on irrelevant terms in $\eps_{\bk}$ is obtained from the 
random phase approximation 
\begin{equation}\label{eq5}
 \Sg^{RPA}(\bk,\xi) = 
 i \int {d^2 q \over (2\pi)^2} \int {d\om \over 2\pi}
 \, D(\bq,\om) \, G^{(0)}(\bk\!-\!\bq,\xi\!-\!\om) ,
\end{equation}
where $G^{(0)}$ is the non-interacting Green function and $D$\/ the 
RPA effective interaction between particles with parallel spin projection.
For the locally interacting model, (\ref{eq1}), 
\begin{equation}\label{eq6}
 D(\bq,\om) = {g^2 \Pi^{(0)}(\bq,\om) \over 
 1 - g^2 \left[\Pi^{(0)}(\bq,\om)\right]^2} ,
\end{equation} 
where $\Pi^{(0)}$ is the non-interacting polarization function.
Numerical results for $\Im\Sg^{RPA}$ are shown in Fig.\ \ref{fig3}.
In contrast to the single sharp peak near $\xi = \xi_{\bk}$ obtained in 
perturbation theory and TMA, the RPA result exhibits two smooth maxima 
with a width that is of the same order of magnitude as their height.
These maxima are due to long-wavelength collective charge and spin density
fluctuations which contribute significantly to the RPA effective 
interaction; such collective fluctuations are not described by perturbation
theory or TMA. 
The corresponding spectral function has Lorentzian shape, with a smooth 
background that vanishes quickly in the low energy limit. 
The quasi-particle decay rate $\gam_{\bk}$ can thus be computed
unambiguously from $\Im\Sg^{RPA}(\bk,\xi_{\bk})$ and turns out to be
proportional to $(k\!-\!k_F)^2 \log|k\!-\!k_F|$ for arbitrary $\eps_{\bk}$
(while the perturbative $\Im\Sg^{(2)}(\bk,\xi_\bk)$ is infinite for a 
linear $\eps_{\bk}$, and finite else). 
The renormalization function $Z(\bk,\xi)$ computed from the RPA
self-energy has a unique limit $Z$ for $\bk \to \bk_F$, $\xi \to 0$.
\smallskip


The peaks in $\Sg(\bk,\xi)$ near $\xi = \xi_{\bk}$ are produced by
virtual excitations where the excited particles and holes have low
energies and momenta near the Fermi surface close to the quasi-particle
momentum $\bk$.
In 3D systems the phase space for these particular excitations is very
small and the low-energy behavior of $\Im\Sg$ is dominated by low-energy
particle-hole excitations with momenta all over the Fermi surface. 
Second order perturbation theory, TMA and RPA all yield the same simple 
behavior $\Im\Sg(\bk,\xi) \propto \xi^2$ with a possible weak 
$\bk$-dependence in the prefactor. 
In fact a renormalized perturbation theory where the bare interaction
is replaced by the exact quasi-particle scattering amplitudes
yields the exact leading low-energy terms of $\Im\Sg$ in 3D already
at second order (in the scattering amplitudes) \cite{SR}.
For dimensionality $d \leq 2$ this is not true because virtual excitations
with {\em small momentum transfers}\/ dominate $\Im\Sg$ \cite{MCD}. 
Their contribution depends on the strength of the interaction vertex for
small momentum transfers $\bq$ (i.e.\ in the forward scattering channel)
which is itself strongly renormalized by low-energy excitations \cite{MCD}!
Since the polarization function $\Pi^{(0)}(\bq,\om)$ has contributions
exclusively from low-energy states for small $\bq$, it is reasonable to
reorganize the perturbation series by replacing bare interactions with RPA
effective interactions. 
The simplest approximation for the self-energy is then to dress the
bare fermionic propagator by only one effective interaction.
It has been pointed out already earlier by Eschrig et al. \cite{EHR}
that all RPA diagrams yield equally important contributions to $\Sg$ in
a renormalized low-energy expansion in two dimensions.

The random phase approximation is not an arbitrary resummation of Feynman
diagrams, but is controlled by the expansion parameter $1/N$, where $N$
is the number of spin states in a generalization of model (\ref{eq1}), where 
instead of two possible spin-projections one allows for $N$ different spin
states. Rescaling the coupling constant by a factor $1/N$, one can take
the large $N$ limit and classify Feynman diagrams by powers of $1/N$. 
Since each interaction line in a diagram yields a factor $1/N$ and
each closed loop a factor $N$, all non-RPA contributions are of order 
$(1/N)^2$. In contrast to plain perturbation theory and the TMA, which
are controlled by the coupling constant $g$ and factors $1/|\log(n)|$,
respectively, the RPA yields a {\em regular}\/ result for $\Sg(\bk,\xi)$.
Thus $1/N$ seems to be a suitable expansion parameter for 2D Fermi liquids,
with higher order corrections being {\em uniformly}\/ small for all $\bk$,
$\xi$ (for large $N$).

\begin{figure}
\epsfbox{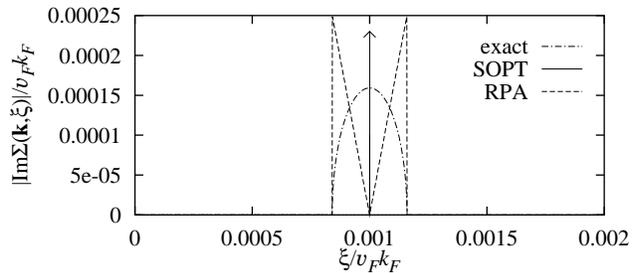}
\caption{$\Im\Sg(k,\xi)$ at fixed $k = 1.001 k_F$ for a
 one-dimensional model with a linear dispersion relation and a coupling
 $g = 1 v_F$; the exact result is compared to perturbation theory and 
 RPA (the vertical arrow represents a $\delta$-function).}
\label{fig4}
\end{figure}

Even more convincing is a comparison with one-dimensional Fermi systems, 
where exact results are available \cite{VOI}. 
Since the peaks in the self-energy discussed above are due to excitations
of particles and holes with almost equal momenta, it is sufficient to
consider a special case of the exactly soluble Luttinger model where 
only the particles and holes near the right (or left) Fermi point interact.
In Fig.\ \ref{fig4}
we show the exact result for $\Im\Sg(k,\xi)$ for this model,
compared to the RPA result and the result of second order perturbation
theory. The sharp peak (a $\delta$-function for a linearized dispersion) 
appearing in perturbation theory is replaced by a smooth maximum with equal
height and width in the exact solution. The RPA correctly captures the
width and height of the exact function, though not its form. One may thus
guess that the two broad maxima in $\Im\Sg(\bk,\xi)$ obtained from the
RPA in 2D have to be replaced by one broad maximum in the full solution. 
In any case the spectral function will have Lorentzian shape.
Note that the RPA correctly signals all the non-Fermi liquid singularities
in one dimension ($\Im\Sg^{RPA}(k,\xi)$ is linear in $k\!-\!k_F$, $\xi$ 
in 1D), while it yields regular Fermi liquid behavior in 2D.

\smallskip
Our explicit (numerical and analytical) results have been obtained for
an isotropic model with a weak local interaction. 
The qualitative conclusions hold more generally for any Fermi
liquid with short-range interactions and a Fermi surface with finite 
curvature, as long as no instability associated with diverging
renomalized interactions occurs.
\medskip

\section*{\normalsize{4.~Conclusion}}

In summary, we have shown that the spectral function for single-particle
excitations in a two-dimensional Fermi liquid with short-range 
interactions has Lorentzian shape. 
Quasi-particles have a uniquely defined spectral weight $Z$ (equal to
the jump of the momentum distribution across the Fermi surface) and a
decay rate proportional to $k'^2 \log k'$, where $k'$ is the distance 
of the quasi-particle momentum from the Fermi surface.
The latter result is not new, but is has so far been derived only by
"good luck", i.e.\ by computing $\Im\Sg(\bk,\xi_{\bk})$ within 
perturbation theory \cite{HSW} or TMA \cite{BLO} for a quadratic dispersion,
where the spurious peak in $\Im\Sg(\bk,\xi)$ reaches its maximum at an
energy of order $\xi_{\bk}^2/\eps_F$ {\em below}\/ $\xi_{\bk}$! 
We have provided several arguments showing that the random phase 
approximation yields a qualitatively correct result for the spectral 
function, while plain perturbation theory and the T-matrix approximation
produce spurious singularities.  

%
\vspace*{2ex}
\noindent{\itshape Acknowledgements:}
We are grateful to H. Kn\"orrer, E. M\"uller-Hartmann, M. Salmhofer,
and E. Trubo\-witz for valuable discussions. 
This work has been supported by the Deutsche Forschungsgemeinschaft under
contract no. \mbox{Me~1255/4-1}.
%
%
\vspace*{2ex}

\setcounter{equation}{0}
\renewcommand{\theequation}{A.\arabic{equation}}
\section*{\normalsize{Appendix:\\ Self-energy in second order
perturbation theory}}
For the continuum model (\ref{eq1}) the second order contribution to the
(time-ordered) self-energy can be calculated using the two-particle
propagator \mbox{$K(\bp,\omega) =
i\int\frac{d^2k}{(2\pi)^2}\int\frac{d\xi}{2\pi}\, 
G^{(0)}(\bp-\bk, \omega-\xi) \,G^{(0)}(\bk,\xi)$}. In the following
we make use of the spectral representations of the self-energy,
$\Sigma(\bk,\xi) = \int d\xi'\,S(\bk,\xi')/(\xi-\xi'+i0^+\sgn\xi')$,
the free one-particle propagator, 
\parbox{\linewidth}{$G^{(0)}(\bk,\xi) = \int
d\xi'\,\rho^{(0)}(\bk,\xi')/(\xi-\xi'+i0^+\sgn\xi')$, 
and the} two-particle propagator, $K(\bp,\omega) = \int d\omega'\,
\Delta^{pp}(\bp,\omega')/(\omega-\omega'+i0^+\sgn\omega')$. For a
quadratic dispersion
$\epsilon_{\bk}=\bk^2/2m$, $\Delta^{pp}$ is given by \cite{FHN}
\begin{equation}\label{app:pp}
\Delta^{pp}(p,\omega) = \left\{\begin{array}{lll}
0 &\!\!\mbox{if}&\!\!\omega<\omega_0\\
\frac{N_F}{2}\sgn\omega &\!\!\mbox{if}&\!\! \omega_0 < \omega < \omega_-\\
\frac{N_F}{\pi}\arcsin\left(\frac{\omega}{p}\frac{1}{\sqrt{\omega-\omega_0}}
\right)
&\!\!\mbox{if}&\!\!\omega_-<\omega<\omega_+ \\
\frac{N_F}{2} &\!\!\mbox{if}&\!\! \omega_+<\omega ,
\end{array} \right.
\end{equation}
where $\omega_0 := p^2/4-1$, $\omega_\pm := p^2/2\pm p$, and the free
density of states $N_F=k_F/2\pi v_F$. Here and in
the following momentum variables are measured in units of $k_F$
and energy variables in units of $v_Fk_F$.

The second order contribution to the spectral function of the
self-energy can be written as
\begin{equation}\begin{array}{l}\displaystyle
S(\bk,\xi) = g^2\int_0^{\xi}\!\frac{d\omega}{2\pi}
\int\!\frac{d^2p}{(2\pi)^2} \Delta^{pp}(\bp,\omega)\,
\rho^{(0)}(\bp\!-\!\bk,\omega\!-\!\xi)\\ \displaystyle
\quad = \frac{g^2N_F}{2\pi}v_Fk_F \int_0^\xi\! d\omega
\int_0^{\infty}\! dp\,p\, \Delta^{pp}(p,\omega)
\bar{\rho}(k,p;\omega-\xi) .
\end{array}\end{equation}
Here 
\begin{equation}\label{app:rhobar}\begin{array}{l}\displaystyle
\bar{\rho}(k,p;\zeta) := 2\int_0^\pi d\phi\, \delta(\zeta-
\xi_{\bp-\bk})\\ \displaystyle
\quad= 4\frac{\Theta\left(
(p^2-(k-\sqrt{2\zeta+1})^2)((k+\sqrt{2\zeta+1})^2-p^2)\right)}{
\sqrt{(p^2-(k-\sqrt{2\zeta+1})^2)((k+\sqrt{2\zeta+1})^2-p^2) }} 
\end{array}\end{equation}
is the angle-integrated imaginary part of the free propagator
$G^{(0)}$, $\phi$ being the angle between $\bp$ and $\bk$ \cite{FHN}.

\begin{figure}
\epsfbox{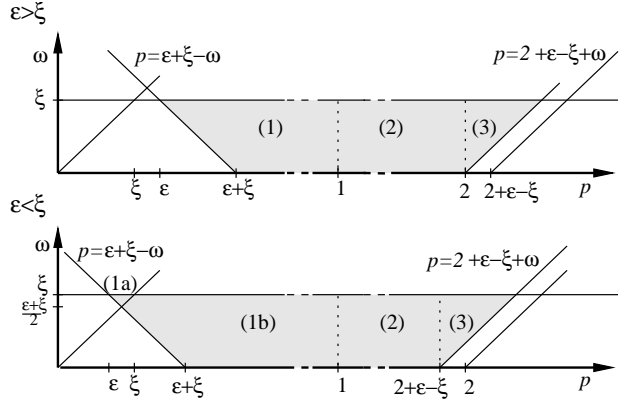}
\caption{Integration boundaries in the $(p,\omega)$ plane for
$0\le\xi\ll 1$ and $\epsilon:=k-k_F\ge0$. $p$, $\epsilon$ in units of
$k_F$; $\omega$, $\xi$ in units of $v_Fk_F$.} 
\label{fig5} 
\end{figure}

For small $\xi\ll 1$ the integration boundaries
$|k-\sqrt{2\zeta+1}|<p<k+\sqrt{2\zeta+1}$, given by the step-function
$\Theta$ in eq.\ (\ref{app:rhobar}), can be linearized to $|k-\zeta-1|
< p<k+\zeta+1$. The same can be done with the boundaries $\omega_0$,
$\omega_\pm$ in eq.\ (\ref{app:pp}). This leaves us with trapezoid
integration regions as given by the shaded areas in Fig.\
\ref{fig5} for the case $\xi\ge0$ and $k-k_F\ge0$. 
Also shown in Fig.\ \ref{fig5} are subareas $(1)$ -- $(3)$, which
can be defined analogously for other choices of $\xi$ and $k$. In
these subareas we can use the following asymptotic expressions for
$\bar{\rho}$ and $\Delta^{pp}$ ($\epsilon := k-k_F$):
\begin{equation}
\bar{\rho}(k,p,\zeta) \approx \left\{\! \begin{array}{lll}
\frac{2}{\sqrt{p^2-(\epsilon-\zeta)^2}} &\! \mbox{if} &\!
(p,\zeta)\in(1),(1a),1(b)\\
\frac{4}{p\sqrt{(2+\epsilon+\zeta)^2-p^2}}&\! \mbox{if} &\!
(p,\zeta)\in(2)\\
\frac{2}{\sqrt{(2+\epsilon+z)^2-p^2}}&\! \mbox{if} &\!
(p,\zeta)\in(3) ,
\end{array}\right.
\end{equation}
\begin{equation}
\Delta^{pp}(p,\omega) \approx \left\{\!\!\begin{array}{lll}
\frac{N_F}{2}\sgn \omega &\!\!\!\!\mbox{if}&\!\!\!\! (p,\zeta)\in(1a)\\
\frac{N_F}{\pi}\arcsin\frac{\omega}{p} &\!\!\!\!\mbox{if}&\!\!\!\!
(p,\zeta)\in(1),(1b)\\
\frac{N_F}{\pi}\frac{2\omega}{p\sqrt{(2+\omega)^2-p^2}}&\!\!\!\!\mbox{if}
&\!\!\!\! (p,\zeta)\in(2) \\ 
\frac{N_F}{\pi}\arctan\frac{\omega}{\sqrt{(2+\omega)^2-p^2}} &
\!\!\!\!\mbox{if} &\!\!\!\! (p,\zeta)\in(3) .
\end{array}\right.
\end{equation}
These approximations lead to a result for the self-energy which is
exact within second order in $\xi$ at fixed $\kappa:=\xi_\bk/\xi \neq
1$. 

With these asymptotic expressions the integration over $p$ can be
performed analytically. The results for the different subareas are
rather lengthy, but can be simplified by a further expansion in
$\omega=\cal{O}(\xi)$ and $\epsilon=\cal{O}(\xi)$, keeping the
ratio $\kappa=\epsilon/\omega$ fixed. Now also the integration over
$\omega$ can be performed analytically, resulting in the simple
expression eq.\ (\ref{eq3}) (Note that $S(\bk,\xi)=
|\Im\Sigma(\bk,\xi)|/\pi$). The main contributions to $S$ come from
the momenta $p\approx 0$ and $p\approx 2k_F$. The contributions
from $p\approx0$ are 
\begin{equation}
\begin{array}{l}\displaystyle
S_{p\approx 0}(\bk,\xi) = g^2\frac{N_F^2v_Fk_F}{2\pi^2}\xi^2
\left[ 1+\ln 2-\ln|\xi|  -\frac{1}{2}\kappa - \phantom{\frac{1}{2}}
\right.\\ \displaystyle
\qquad\qquad\qquad\left. -\ln\left| \kappa+1\right|
+\frac{1}{4}\left(\kappa-1\right)^2 
\ln\left|\frac{\kappa+1}{\kappa-1}\right| \right] .
\end{array}
\end{equation}
The contributions from $p\approx 2k_F$ are
\begin{equation}
\begin{array}{l}\displaystyle
S_{p\approx 2k_F} (\bk,\xi) = g^2\frac{N_F^2v_Fk_F}{2\pi^2}
\xi^2 \times \\ \displaystyle
\qquad\qquad\qquad\, \times \left[\ln 2\sqrt{3}
-\frac{1}{2}\ln|\xi| -\frac{1}{2}\ln\left|\kappa-1\right|\right] ,
\end{array}
\end{equation}
causing the logarithmic divergence at $\kappa=1$.

\renewcommand{\refname}{\normalsize{References}}

%
%
%
%
%
%
%
%
%

\begin{thebibliography}{99}
\bibitem{EMH} We would like to thank E. M\"uller-Hartmann for drawing
our attention to this problem.
\end{thebibliography}
\end{document}